\documentclass[twocolumn,floatfix,prb,showpacs,superscriptaddress]{revtex4}
\usepackage{graphicx,color,bm}
\usepackage{amsbsy,amssymb,amsmath,bm}

\usepackage{graphicx}
\usepackage{epstopdf}
\usepackage{latexsym}
\usepackage{subfigure}
\usepackage{color}
\usepackage{hyperref}

\newcommand{\ccc}{{Cs$_2$CuCl$_4$}}
\newcommand{\be}{\begin{equation}}
\newcommand{\ee}{\end{equation}}
\newcommand{\bea}{\begin{eqnarray}}
\newcommand{\eea}{\end{eqnarray}}

\begin{document}

\title{High-field magnetic resonance  of spinons and magnons \\ in a
triangular lattice $S=1/2$ antiferromagnet Cs$_{2}$CuCl$_{4}$ }

\author{A.~I.~Smirnov}
\affiliation{P.~L.~Kapitza Institute for Physical Problems, RAS, 119334 Moscow, Russia}

\author{T.~A.~Soldatov}
\affiliation{P.~L.~Kapitza Institute for Physical Problems, RAS, 119334 Moscow, Russia}
\affiliation{{Moscow
Institute for Physics and Technology, 141700 Dolgoprudnyi, Russia}}

\author{{K.~Yu.~Povarov}}
\altaffiliation[Previous address: ]{P. L. Kapitza Institute for
Physical Problems RAS}
\affiliation{Neutron Scattering and
Magnetism, Laboratory for Solid State Physics, ETH Z\"{u}rich,
Switzerland}

\author{A.~Ya.~Shapiro}
\affiliation{A.~V.~Shubnikov Institute of Crystallography, RAS, 119333, Moscow, Russia}

\begin{abstract}
The electron spin resonance doublet indicating the width of the two
spinon continuum in a spin-1/2 triangular-lattice Heisenberg
antiferromagnet Cs$_2$CuCl$_4$ was studied in high magnetic field.
The doublet was found to collapse in a magnetic field of a half of
the saturation field. The collapse of the doublet occurs via
vanishing of the high frequency  component in a qualitative
agreement with the theoretical prediction for the $S=1/2$
chain. The  field of the collapse is, however, much lower than
expected for the $S=1/2$ chain. This is proposed to be due to the
destruction of frustration of interchain exchange bonds in a
magnetic field, which restores the 2D character of this spin system.
In the saturated phase the mode with the Larmor frequency and a much
weaker  mode downshifted for 119~GHz are observed. The weak mode is
of exchange origin, it demonstrates a positive frequency shift at
heating corresponding to the repulsion of magnons in the saturated
phase.

\end{abstract}

\maketitle

\section{Introduction}\label{Introduction}
Spin-1/2 Heisenberg antiferromagnet on a triangular lattice \ccc \ was extensively studied because of a
remarkable two-spinon continuum of excitations, like that of the $S=1/2$ antiferromagnetic
chain\cite{ColdeaPRL, Coldea03}. Other feature of this material is the delayed magnetic ordering at very low
temperature and exotic field-induced phase transitions\cite{Tokiwa}, which are due to weak interactions, while
the dominant exchange interaction is frustrated\cite{Starykh10}. The aim of this work is to study the fine
structure of the spinon continuum in a high magnetic field, including the transition to saturated phase.

\begin{figure}[t]
\begin{center}
\includegraphics[width=0.4\textwidth]{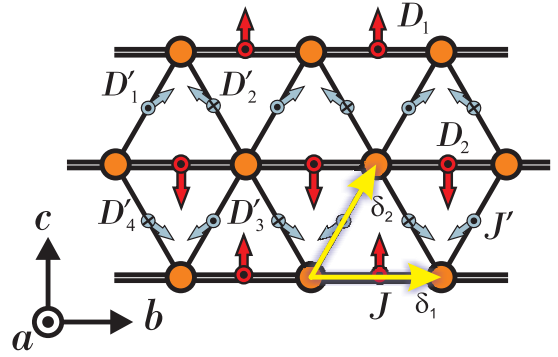}
 \caption{(Color online) Sketch of the exchange paths of \ccc \ within
 a $bc$-layer. Large circles mark Cu-ions.
 $J$ and $J'$ are exchange integrals for two kinds of bonds. ${\bf D}_{1,2}$ and ${\bf D}'_{1-4}$
 are Dzyaloshinsky-Moriya vectors according to Ref.~\onlinecite{Starykh10}.
 Out-of-plane components of vectors are marked by points and crosses.
 Translations $\delta_{1,2}$ are periods for exchange bond
 structure.}\label{fig:J&J'}
\end{center}
\end{figure}

In crystals of \ccc \  magnetic ions Cu$^{2+}$  ($S=1/2$) are displayed in layers with a distorted triangular
lattice, see Fig.~\ref{fig:J&J'}. The 2D model Hamiltonian contains the following terms, see, e.g.,
Ref.~\onlinecite{Starykh10}:

\begin{equation}
\hat{\mathcal{H}} = J\sum_{\langle i,j \rangle} {\bf S}_i\cdot{\bf S}_j +
J' \sum_{\langle i,j' \rangle}  {\bf S}_i\cdot{\bf S}_{j'}+
 \sum_{\langle i,k \rangle}  {\bf D}_{ik}\cdot{\bf S}_i \times {\bf S}_{k} \ , 
\label{Ham}
\end{equation}
here  $J$ is the exchange integral for spins neighboring along $b$-direction,  $J'$ is the zig-zag interchain
coupling, as shown in Fig.~\ref{fig:J&J'}. Vectors ${\bf D}_{ik}$ are  parameters of Dzyaloshinsky-Moriya
interaction. There are six  different Dzyaloshinsky-Moriya vectors (${\bf D}_{1,2}$ and ${\bf D}'_{1-4}$)
compatible with the symmetry of \ccc\cite{Starykh10}. These vectors are shown schematically in Fig.
\ref{fig:J&J'} in the middle of each exchange bond.  Vectors ${\bf D}_{1,2}$  have nonzero $a$- and
$c$-components of absolute values $D_a$ and $D_c$ and are oriented as shown in Fig. \ref{fig:J&J'}. Vectors
${\bf D}'_{1-4}$ have nonzero components along all three crystallographic axes with absolute values
$D'_{a,b,c}$.

The main exchange integrals $J, J'$ and the interplane exchange constant $J''$ were derived from the neutron
scattering experiments in the saturated phase \cite{Coldea02}: $J$=4.34(6)~K, $J'$=1.48(6) ~K, $J''$=0.22(3)~K.
Close values $J$=4.7(2)~K, and $J'$=1.42(7)~K were derived from the saturation field value\cite{Tokiwa} and
electron spin resonance (ESR) in the saturated phase.\cite{ZvyaginPRL2014} Several parameters of
Dzyaloshinsky-Moriya interactions were extracted from experiments:  inelastic neutron scattering\cite{Coldea02}
gives  $D'_a$=0.24~K, low-temperature ESR\cite{Povarov,Smirnov} results in $D_a$=0.23$\pm$0.05~K,
$D_c$=0.34$\pm$0.05~K,  and high-temperature ESR linewidth\cite{Fayzullin} gives $D_{a}=0.33$~K, $D_{c}=
0.36$~K.

The observation of 1D excitation spectrum in this quasi 2D system is ascribed to the effective decoupling of
spin chains because of the frustration of the antiferromagnetic exchange interactions $J'$ on the diagonal
bonds. This effective decoupling is supported by, e.g., numerical simulations\cite{Becca} and analytical
approach\cite{StarykhNature}.

 A feature of this compound is the uniform Dzyaloshinsky-Moriya interaction between the spins,
neighboring along $b$-direction: vectors ${\bf D}_{1,2}$ are equal in magnitude and direction for all bonds
within a chain, in contrast to vectors ${\bf D'}_{1-4}$ on diagonal bonds, which compose a staggered structure
of a conventional Dzyaloshinsky-Moriya interaction. The uniform Dzyaloshinsky-Moriya interaction was predicted
to modify the spectrum of excitations  in a 1D $S$=1/2 chain. This interaction should cause a shift of the
continuum in ${\bf q}$-space by a vector $q_{DM}=\frac{D}{J}\frac{\pi}{b}$. As a result, in a magnetic field
${\bf H}\parallel{\bf D}$,  the ESR line should split into a doublet. The frequencies of the doublet components
are at the upper and lower boundaries of the initial (i.e. unshifted) continuum at the wave vector $q_{DM}$, see
Refs.~\onlinecite{StarykhESR,Affleck2011}.   At the same time, ESR signal should not split at the orthogonal
orientation of magnetic field. In this case a gap of the ESR absorption in zero field should open. The doublet
is marking the width of the continuum and  appears due to the fractionalized character of excitations and to
uniform Dzyaloshinsky-Moriya interaction.

\begin{figure}[t]
\begin{center}
\includegraphics[width=0.5\textwidth]{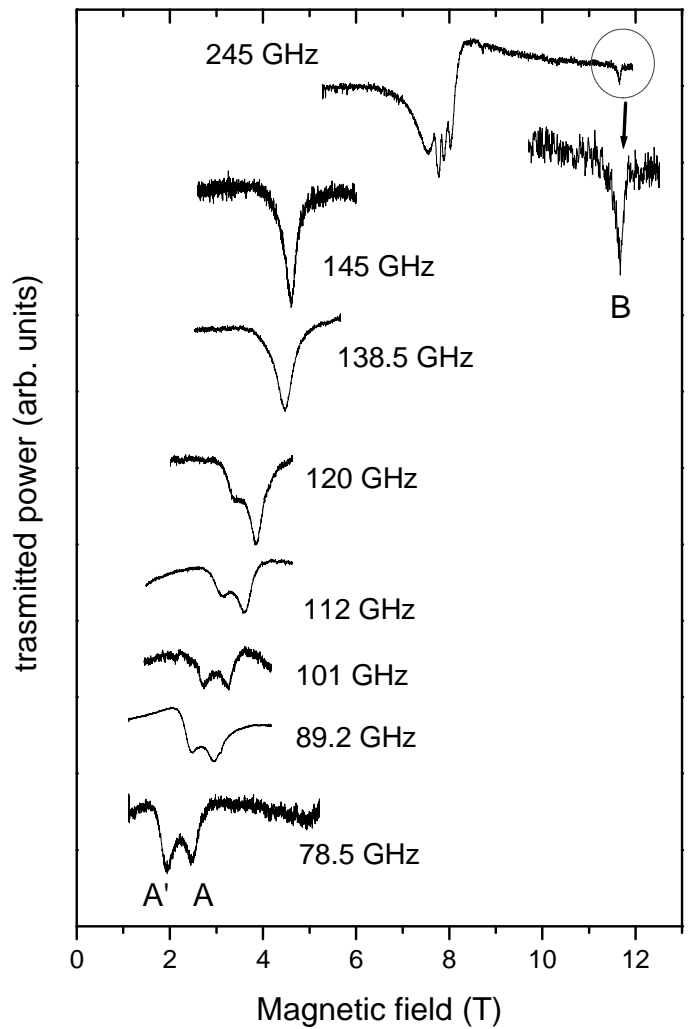}
\caption{\label{fig:ESRlines@varfreq}  Examples of ESR lines of \ccc \  at ${\bf
 H}
\parallel a$  at different frequencies.
The zoom for the area within a circle is 6-fold for vertical scale and 3-fold for horizontal scale.  }
\end{center}
\end{figure}

 The ESR doublet arising at ${\bf H} \parallel {\bf D}$  and merging into a single line at ${\bf H} \perp {\bf D}$
 was indeed observed experimentally in the spin-liquid phase of \ccc \ in
Refs.~\onlinecite{Povarov,Smirnov}.

In the preceding ESR paper describing high-field experiments\cite{ZvyaginPRL2014} magnetic field was oriented
parallel to  $b$-axis, because the exchange parameters may be derived most accurately by use of this
orientation. However, at ${\bf H}
\parallel b$   magnetic field is perpendicular to ${\bf D}$ and the
spinon doublet does not appear. Therefore, the high-field evolution of the ESR spinon doublet (i.e. of the width
of the spinon continuum near the Brillouin zone center) was not addressed there. The aim of the present work is
to follow experimentally the evolution of the ESR doublet in a high magnetic field, i.e at frequencies of the
exchange range. A vanishing of the doublet is expected in a high field, because of the suppressing of quantum
fluctuations. In particular, this should close the width of the spinon continuum at saturation (see theory,
e.g., Ref.~\onlinecite{Mueller}). We indeed detect the vanishing of the doublet at ${\bf H}\parallel a$. It
occurs via the cease of the high frequency component of the doublet. Besides, we observe the transformation of
the spinon ESR response into two ESR modes in the saturated phase. One of these modes originates from the magnon
branch with a dispersion along $c$-axis. This mode shows an effect of the frequency shift at heating, indicating
a repulsion of magnons.

\section{Experiment}\label{Experiment}

The preceding investigation \cite{Smirnov} shows, that the spinon
doublet is most clearly pronounced at lowest temperatures and that
at the frequency above the exchange value  $J/\hbar \simeq 80$~GHz
it is  not affected by cooling below $T_N=0.62$~K. The change of the
doublet to an antiferromagnetic resonance spectrum was observed only
below 40~GHz. This conservation of a spin-liquid spectrum at high
energy range, and arising of spin-wave modes at low energy is
typical for systems with a delayed ordering occurring far below
Curie-Weiss temperature (see, e.g.,
Refs.~\onlinecite{ZaliznyakPetrov,KCuF3}).  For the above reason, to
measure the high-field evolution of ESR signal we choose the
temperature of about 0.5~K and the frequency above 60~GHz. Here the
components of the doublet are well separated and at the same time
are not affected by the ordering. We align the magnetic field along the
$a$-axis, because at this orientation the doublet is well seen and
the spin structure evolves gradually to saturation without phase
transitions even in the ordered phase. We used the crystals of \ccc
\ from the same batch as in
Refs.~\onlinecite{Povarov,Smirnov,ZvyaginPRL2014}.

Experiments  were performed using a set of ESR spectrometers, operating with superconducting 12~T magnet,
combined with a $^3$He cryo-insert, providing low temperature down to~0.45~K. A small amount of powder of
2,2-diphenyl-1-picrylhydrazyl (known as DPPH) was employed as a standard $g=2.00$ marker for the  field.
Backward wave oscillators were microwave sources, covering the range 60--350~GHz. The microwave units of two
types were used for recording the resonance absorption of microwaves. In the first unit cylindrical multi-mode
resonators were used as plug-in components in a transmission microwave circuit. The second unit is a narrowed
waveguide with a diaphragm, also used in a transmission mode. In case of a properly tuned resonator we observe
the diminishing of the transmission, proportional to the imaginary part of the susceptibility of the sample. The
ESR line of a conventional paramagnet recorded in this way should have a Lorentzian shape. Unfortunately, for
frequencies above 200~GHz the spectrum of eigenfrequencies of the cavity is too dense and proper tuning is
difficult. The waveguide doesn't require frequency tuning, but in this case the change of transmission is a
superposition of the real ($\chi'$) and imaginary ($\chi''$) parts of microwave susceptibility (see, e.g.
Ref.~\onlinecite{Gurevich}).

\begin{figure}[h]
\begin{center}
\includegraphics[width=0.5\textwidth]{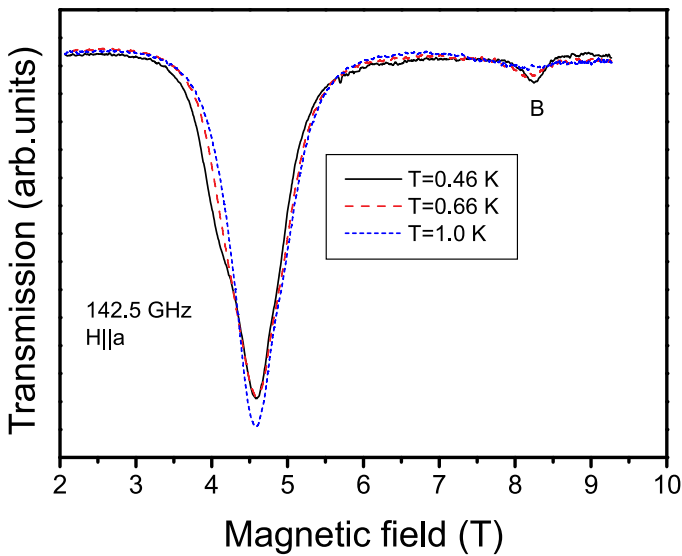}
\caption{\label{modesA&B} (Color online) 142~GHz ESR records at
$\bf{H}\parallel a$ in \ccc \ at several temperatures.}
\end{center}
\end{figure}

It should be noted, that at the frequency above 140~GHz  the samples of a size of about 2 mm have strongly
distorted (indented) ESR line because of parasitic field-dependent resonances, which arise due to the large
dynamic susceptibility $\chi'$ of the sample near the resonance field. The susceptibilities $\chi'$, $\chi''$
change strongly and not monotonously within an interval of several linewidths $\Delta H$ on both sides of the
resonance field $H_0$. For a typical paramagnet  (see, e.g., Ref.~\onlinecite{Abragam}) the susceptibility
$\chi'$ is negative in an interval below the resonance field, then it takes positive value, reaching a maximum
at $H- H_0\simeq\frac{1}{2}\Delta H$, and then gradually drops to zero. Thus, the large positive values of
$\chi'$ occur twice during a sweep of the field across the right wing of the resonance curve. The large value of
$\chi'$ may result in electrodynamic resonances in the dielectric sample at field values, when a half of the
electromagnetic wave length is comparable to the sample size or fractional value of the sample size. In this
way, for a high frequency, large sample and high susceptibility, several electrodynamic resonances may arise in
the field interval, where the real part of the susceptibility is rising and the same resonances should again
occur in the field range, where the susceptibility is falling. In this case the ESR lineshape appears to be
distorted by indenting via parasitic resonances. To avoid this parasitic effect, one has to use a method of
transmission of plane electromagnetic waves through the sample, having a thin plate shape\cite{Mukhin}. Here the
electrodynamic resonances are fixed as interference pattern of plane waves. Another way is to diminish the
sample size far below the half of the length of the electromagnetic wave within the sample. We used the samples
of the size below 0.5~mm for recording strong ESR signals and samples with the size of about 2~mm to detect the
weak ESR line, which arises above the saturation field. Besides, a test for parasitic resonances may be
performed in the paramagnetic phase at $T>10$~K, when the imaginary part of the ESR susceptibility is surely a
Lorentzian function of the magnetic field, and the real part is also a known function of field, see, e.g.,
Ref.~\onlinecite{Abragam}.  The manipulation with the sample size and the test by means of the paramagnetic
resonance enables one to avoid parasitic electrodynamic resonances and to fix the intrinsic lineshape in the
range below 250 GHz. At higher frequencies  ESR lines appeared indented, this resulted in a higher error of the
measurement of resonance field.

\section{Experimental results}\label{Results}

The evolution of the ESR lineshape with changing frequency at ${\bf
H }\parallel a$ is shown in Fig. \ref{fig:ESRlines@varfreq}. Here
the ESR records taken by resonator unit in the range 70--150~GHz and
by waveguide unit at 245~GHz are shown.  We see, that the low-field
(i.e. high-frequency) component of the doublet looses intensity with
the increase of the magnetic field. For frequencies above 140~GHz
the doublet disappears completely, and only a single ESR line with
the paramagnetic resonance frequency $f_0 = g_a\mu_B H/(2\pi\hbar)$,
$g_{a}=2.20$ is observed below the saturation field 8.44~T. At the
further increase of the magnetic field we continue to observe a
strong ESR line with the frequency $f_0 = g_a \mu_B  H/(2 \pi
\hbar)$. Besides, we see a much weaker ESR line in the magnetic
field above 8~T (mode "B"), see upper curve in Fig.
\ref{fig:ESRlines@varfreq}. The ratio of integral intensity of the
weaker mode B to the intensity of the $f_0$ mode is about 0.015. The
245~GHz curve presents an example of the parasitic indention of the
intensive resonance for a sample, which is oversized for the high
microwave susceptibility of $f_0$-mode, but provides a good
sensitivity for a weak mode B.

Fig.~\ref{modesA&B} \ demonstrates records of 142~GHz ESR at several
temperatures. This records, performed by a resonator unit in the
middle part of the frequency range present both intensive and weak
modes for the same sample without parasitic distortions. The
temperature evolution of the intensive line shows vanishing of the
doublet component A$^\prime$ in the almost collapsed doublet by
heating. The weak line appearing above the saturation field also
disappears at heating.

\begin{figure}[h]
\begin{center}
\includegraphics[width=0.5\textwidth]{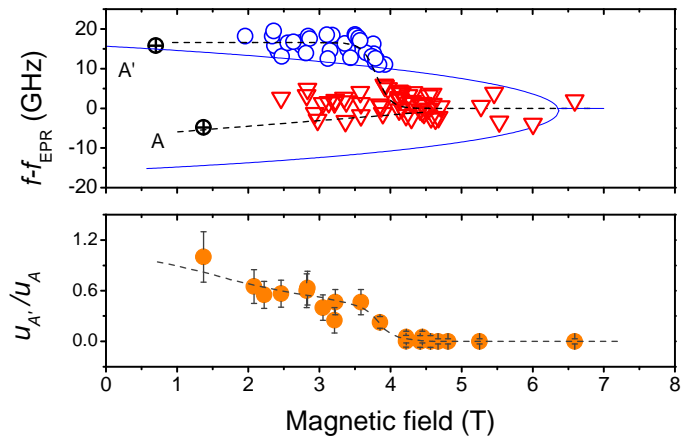}
\caption{\label{fig:doublet-collapse} (Color online) Upper panel: shift of the resonance frequencies of doublet
components $A$ and $A'$ relative to paramagnetic resonance. Crossed circles are 35~GHz data\cite{Smirnov} for
the spin-liquid phase at $T=1.0$~K. Other symbols correspond to T=0.5 K. Solid line presents calculation of the
continuum boundaries at the wavevector $q_D$ following Ref.~\onlinecite{Mueller}. Lower panel: ratio of
amplitudes of the doublet components $u_{A'}/u_{A}$ {\it vs} resonance  field of $A$-component. Dashed lines in
both panels are guides to the eyes. }
\end{center}
\end{figure}

\begin{figure}[h]
\begin{center}
\includegraphics[width=0.5\textwidth]{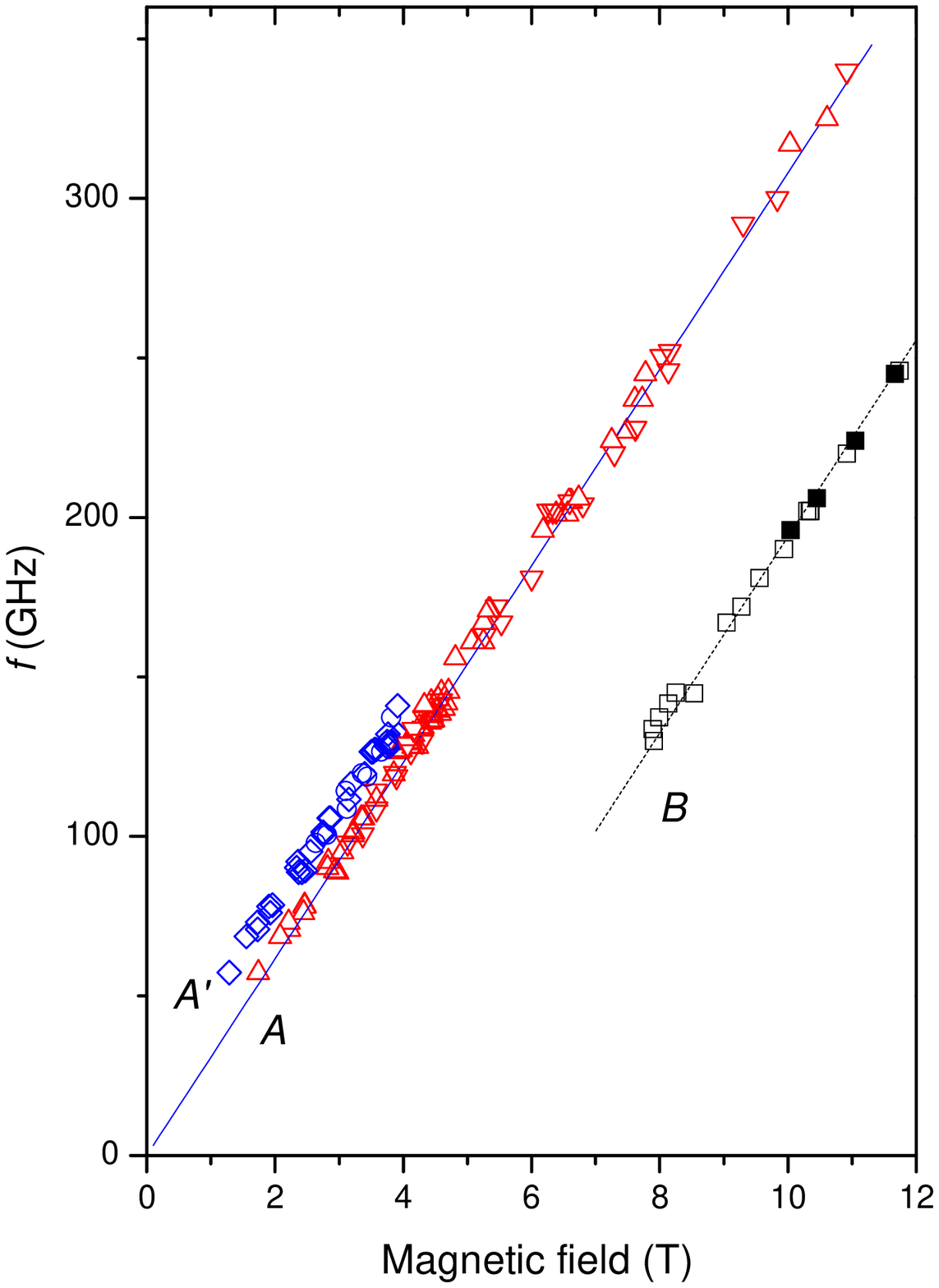}
\caption{\label{fig:ffd-overview} (Color online) Frequency-field diagram of ESR response  in \ccc \ with
magnetic field along the $a$-axis ($T=0.5$~K). Data presented by symbols $\vartriangle$, $\lozenge$,
$\blacksquare$ are taken by use of the waveguide unit, $\triangledown$,$\bigcirc$, $\square$  - by resonator
unit.}
\end{center}
\end{figure}

The transformation of the doublet into a single ESR line with the
increase of the magnetic field  is illustrated in
Fig.~\ref{fig:doublet-collapse}, here the field dependence of the
shift of doublet components with respect to $f_0$ is shown in the
upper panel.  Data presented here  are taken in the frequency range
60--200~GHz at $T=0.5$~K. The lower panel shows the ratio of
amplitude of the upper component $u_{A\prime}$ to the lower component
amplitude $u_{A}$. The collapse of the doublet occurs in the
magnetic field of 4.0~T, which constitutes approximetely $0.5 H_{sat}$.
The frequency-field dependence of all modes at $T=0.5$~K is
presented in Fig.~\ref{fig:ffd-overview}.

\begin{figure}[h]
\begin{center}
\includegraphics[width=0.5\textwidth]{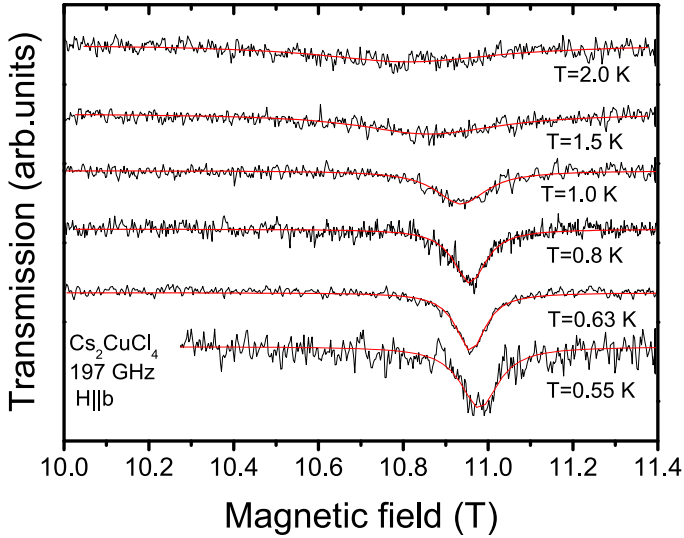}
\caption{\label{lineB-T-evol} (Color online) Temperature evolution
of line B at the frequency 197~GHz and ${\bf H}\parallel b$. Solid
lines are Lorenz-fits for experimental curves.}
\end{center}
\end{figure}

\begin{figure}[h]
\begin{center}
\includegraphics[width=0.5\textwidth]{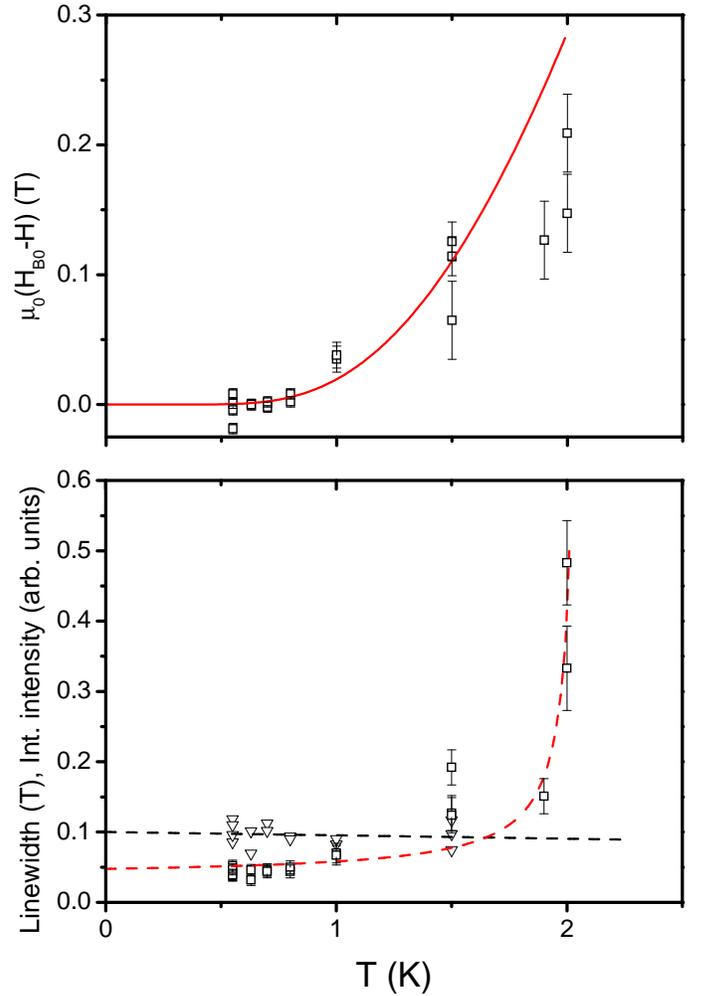}
\caption{\label{T-ind-shift} (Color online) Upper panel: Temperature
induced shift of the resonance fields of line B at the frequency
197~GHz, ${\bf H}\parallel b$. The reference field
$\mu_0H_{B0}$=11~T is the ESR field for mode B at $T=0.5$~K. Solid
line presents the theoretical calculation according to relation
(\ref{deltaepsilon}). Lower panel: Temperature dependence of the
integral intensity (triangles) and linewidth (squares) of mode "B"
at 197~GHz. Dashed lines are guide to the eyes. }
\end{center}
\end{figure}

The weaker mode B, arising above the saturation field, was observed also at ${\bf H} \parallel b$ in
Ref.~\onlinecite{ZvyaginPRL2014} We study here the temperature dependence of this mode. The orientation of the
field ${\bf H} \parallel b$ is selected because the theory\cite{ZvyaginPRL2014} has maximal accuracy  at this
direction of the field. The resonance field of the mode B exhibits a shift to lower fields at heating, as
demonstrated in Fig.~\ref{lineB-T-evol}. The temperature dependence of the resonance field and of the linewidth
are shown in Fig.~\ref{T-ind-shift}.

\section{Discussion}

The spinon continuum and the related ESR doublet observed in \ccc \ are consequences of quantum fluctuations in
a spin system, which remains paramagnetic (spin-liquid) at temperatures far below the Curie-Weiss temperature.
Magnetization should suppress zero-point fluctuations and, hence, the doublet should be transformed in a single
ESR line with the Larmor frequency, at least in the saturated phase. The application (see
Appendix~\ref{AppendixA}) of the theory of a Heisenberg $S$=1/2 antiferromagnetic chain\cite{Mueller} predicts a
collapse of the spinon continuum width (and, hence, of the considered doublet) at the saturation field, see
solid line in the upper panel of Fig.~\ref{fig:doublet-collapse}. The exchange integral $J$ and
Dzyaloshinsky-Moriya parameter $D_a$ of \ccc\ was used for this calculation.
Note, that the saturation field 6.3~T calculated  in 1D model is lower than the more realistic 2D value of 8.44~T.

Our observations confirm that the collapse of the doublet really
occurs. However, the doublet does not survive till the saturation
field, but collapses in the field of about  $0.5H_{sat}$. This
discrepancy between the observed behavior of the spinon doublet and
the theory of a spin chain may be attributed, probably, to the cease
of the frustration of the exchange coupling between chains in \ccc \
in a magnetic field.  The frustration of interchange coupling takes
place in zero field, when the antiferromagnetic correlation of
neighboring spins within the chain prevails\cite{Becca}. The
antiferromagnetic correlation changes to a ferromagnetic one in a
strong magnetic field, thus, the interaction between the chains
should be restored and 1D consideration  becomes inapplicable to
\ccc \ in a strong field.

The vanishing of the upper component of the doublet is qualitatively
consistent with the theoretical investigation of the spectral
density of two spinon continuum of the 1D $S=1/2$ Heisenberg
antiferromagnet in a magnetic field\cite{Kohno}. In this theoretical
study the intensity at the upper boundary of the spinon continuum of
the transverse spin oscillations is shown to drop in process of
magnetization (see Fig. 3 of Ref.~\onlinecite{Kohno}).

The experiment in the field, exceeding the saturation point $\mu_0 H^a_{sat}$ = 8.44~T, shows the downshift
$119\pm2$~GHz of the frequency of the weak mode B with respect to the Larmor frequency of mode A. The appearance
of a downshifted mode  is in a good correspondence with the theory of the spin wave excitations of the saturated
phase, given in the Supplement material of Ref.~\onlinecite{ZvyaginPRL2014}. This theory predicts approximately
the same ESR frequencies for the field both perpendicular and tilted to vector ${\bf D}$:

  \begin{equation}
 2\pi\hbar f_{ESR} = g_{a,b,c}\mu_B H_{a,b,c} - 4J'+ O(D/J)^2.
 \end{equation}

   The same shift 119~GHz was observed for the mode B for another  orientation of the magnetic field
   ${\bf H} \parallel b$ in Ref.~\onlinecite{ZvyaginPRL2014}.  Thus, the theoretical prediction on the approximately
   isotropic character
   of the shift of mode B corresponds
   well to the theory. It should be noted, that the mode B, observed here by ESR method, i.e. at the
   Brillouin zone center, is the same excitation, as observed by inelastic neutron scattering\cite{Coldea02} at the
   boundary of the "exchange" Brillouin zone ($k_c=\pi/c$). Indeed, in  Ref.~\onlinecite{Coldea02} the Brilloun zone
   was considered  in
   the exchange approximation with periods  $\delta_{1,2}$. However, the
   structure composed by vectors ${\bf D}'$ has a doubled
   period in $c$-direction in comparison with the exchange structure, see
   Fig.~\ref{fig:J&J'}.  The period
   doubling results in the folding of the Brilloun zone. Thus,  excitations, positioned at
   the boundary of the zone in the exchange
   approximation,  appear in the center of the zone (see Fig. 1 in Ref.~\onlinecite{ZvyaginPRL2014}).
   The weak but nonzero ESR intensity of this mode is attributed, thus, to the
   Dzyaloshinsky-Moriya interaction and would be zero in the exchange approximation (see theory in
    Ref.~\onlinecite{ZvyaginPRL2014}).

 The negative shift of the resonance field of mode B, observed at heating, means the enlarging of the eigen frequency
of magnons  at excitation of additional magnons. This may be treated
as a consequense of a repulsive interaction of  magnons.  The
repulsion of magnons is natural for fully polarized
antiferromagnetic system, because here two flipped spins show a
mutual  repulsion. The  mode B practically disappears above 2~K.
This is also natural as the dispersion in $k_c$-direction is
provided by the exchange $J'=1.45$~K. This dispersion determines the
frequency of the ESR mode B.  Thus, the  temperature, higher than
$J'$, should smear this resonance mode, as seen in the experiment.

The nonlinear spin-wave calculations within $J$-$J'$ Heisenberg
model in the saturated phase \footnote{M. Zhitomirsky (private
communication)}\ give the following expression for the
temperature-dependent energy shift at the wavevector $q=(0,
2\pi/3)$, corresponding to the frequency of ESR mode B:

\begin{equation}
\delta \varepsilon = 8J'\sum_{{\bf k}}
\frac{1-\cos\frac{k_x}{2}\cos\frac{\sqrt{3}k_y}{2}}{\exp(\frac{\varepsilon({\bf k})}{k_BT})+1}
\label{deltaepsilon}
\end{equation}

   Here the wavevector is measured in units of reciprocal periods of the 2D lattice with the translations
$\delta_{1,2}$ in Fig.~\ref{fig:J&J'}. The magnon spectrum
$\varepsilon({\bf k})$ is described by the relation 26 of
Supplemental material of Ref.~\onlinecite{ZvyaginPRL2014}.

The result of the calculation  after  relation (\ref{deltaepsilon}) is shown in the upper panel of
Fig.~\ref{T-ind-shift}. This calculation is made for $\mu_0 H=11$~T and exchange parameters $J$ and $J'$ for
\ccc \ from Ref.~\onlinecite{ZvyaginPRL2014}. The result of the experiment corresponds well to the theory both
in the sign and the value of the shift.

\section{Conclusion}
   The evolution of the electron spin resonance spectrum in the frequency range above the exchange frequency
   $J/(2\pi\hbar)$ was studied in the $S=1/2$ antiferromagnet on the distorted triangular lattice
   \ccc. The doublet of resonance lines, marking the boundaries of the spinon continuum was found to collapse in
   the field of about a half of the saturation field. The collapse proceeds via vanishing of the upper
   frequency component of the doublet. This scenario of the collapse of the doublet agrees  qualitatively with the
   evolution of the spinon continuum of spin $S=1/2$ Heisenberg antiferromagnetic chain
    \cite{Mueller,Kohno}. Above the saturation field,
   a much weaker mode, downshifted for 119~GHz from the Larmor frequency is observed.
   This shift and the weak intensity of this mode
   correspond well to the theoretical consideration
   of  spin waves in the saturated phase.
   The temperature dependence of the resonance field of the weaker
   mode indicates the repulsive interaction of magnons in the saturated antiferromagnet and is well explained
   within the spin wave formalism with anharmonic terms.

  We thank M.E. Zhitomirsky for the calculation of temperature dependent shift of the
magnon spectrum in the saturated phase and for valuable discussions,
A.I. Kleev for the analysis of the transmission through the
waveguide with a magnetic sample,  V.N.~Glazkov, S.S.~Sosin,
O.A.~Starykh, L.E. Svistov for numerous discussions and comments.
Work at the Kapitza Institute is supported by Russian foundation for
basic research, grants No. 12-02-00557, 15-02-05918.\\

\appendix
\section{ESR frequencies of the spin chain in high fields\label{AppendixA}}

Under the action of the applied magnetic field continua of spin
fluctuations of different polarization become different.  The
spectra of continua of transverse spin fluctuations $S_{+-}$ and
$S_{-+}$  are responsible for ESR absorption\cite{OshikawaAffleck}.
The upper and lower boundaries of these continua for $S=1/2$
Heisenberg antiferromagnetic chain in presence of a magnetic field
are calculated by M\"{u}ller \emph{et al.} These data are summarized
in Table~II of Ref.~\onlinecite{Mueller}. As described in Sec. I, to
calculate the upper and lower boundary frequencies at $q=0$ for a
spin chain with the uniform Dzyaloshinsky-Moriya interaction we use
the results of Ref.~\onlinecite{Mueller}, taking the boundary
frequencies at  $q=q_{DM}$ for the boundaries with non-vanishing
spectral weight. In this way we obtain the following frequency-field
dependencies for spin-resonance absorption at $\mathbf{H}\parallel
\mathbf{D}$:

\begin{equation}\label{EQ:ESRmode1}
    2\pi\hbar f_{1}=JR(h)\sin\left(\frac{q_{\text{DM}}}{2}\right)\cos\left(\frac{q_{\text{DM}}}{2}-\pi
    m(h)\right)-Jh,
\end{equation}

\begin{equation}\label{EQ:ESRmode2}
    2\pi\hbar f_{2}=JR(h)\sin\left(\frac{\pi}{2}-\frac{q_{\text{DM}}}{2}\right)\cos\left(\frac{\pi}{2}-\frac{q_{\text{DM}}}{2}-\pi
    m(h)\right),
\end{equation}

and

\begin{equation}\label{EQ:ESRmode3}
    2\pi\hbar f_{3}=JR(h)\sin\left(\frac{\pi}{2}+\frac{q_{\text{DM}}}{2}\right)\cos\left(\frac{\pi}{2}+\frac{q_{\text{DM}}}{2}-\pi
    m(h)\right).
\end{equation}

Here $h=g\mu_{\text{B}}\mu_{0}H/J$ is the reduced field
($h_{sat}=2$), $R(h)=\left(\pi+h(1-\frac{\pi}{2})\right)$ is the
field-dependent renormalization prefactor and $m(h)$ is the reduced
magnetization given by\cite{Mueller}

\begin{equation}\label{EQ:ReducedM}
    m(h)=\frac{1}{\pi}\arcsin\left(\frac{1}{1-\pi/2+\pi/h}\right).
\end{equation}

Domains of these functions in $q$ and $H$ are chosen to avoid
negative values of frequencies. Using equations (\ref{EQ:ESRmode1}
--- \ref{EQ:ReducedM}) we get the ESR frequencies  shown in
Fig.~\ref{fig:doublet-collapse} for the $S=1/2$ spin chain with
uniform Dzyaloshinsky-Moriya interaction in the so-called
\emph{M\"{u}ller ansatz} approximation: mode $f_{1}$ is given by the
lower boundary of $S_{-+}$ continuum and $f_{2,3}$ are given by the
lower boundary of $S_{+-}$ continuum. In case of low field $h\ll1$
these equations transform into corresponding relations of
Refs.~\onlinecite{StarykhESR,Povarov}. Modes $f_2$ and $f_3$
correspond to A$'$ and A in Fig.~\ref{fig:doublet-collapse}, and
mode $f_1$ is relevant only for small magnetic fields which are out
of range of the present study.

\bibliography{BibfileESRcollapse}
\bibliographystyle{apsrev4-1}

\end{document}